\documentclass[twocolumn,pra,superscriptaddress]{revtex4-1}
\usepackage{graphicx}
\usepackage{epsfig}
\usepackage{ulem}
\usepackage{color}
\usepackage{braket}
\usepackage{amsmath,amsfonts,amsthm,bm}
\usepackage[breaklinks,colorlinks,linkcolor=blue,citecolor=blue,urlcolor=blue]{hyperref}

\begin{document}

\title{Quantum Dynamics of Cold Atomic Gas with $SU(1,1)$ Symmetry}

\author{Jing Zhang}
\affiliation{School of Physics, Xi'an Jiaotong University, Xi'an 710049, Shaanxi, China}
\affiliation{Shaanxi Province Key Laboratory of Quantum Information and Quantum Optoelectronic Devices, Xi'an Jiaotong University, Xi'an 710049, Shaanxi, China}

\author{Xiaoyi Yang}
\affiliation{School of Physics, Xi'an Jiaotong University, Xi'an 710049, Shaanxi, China}
\affiliation{Shaanxi Province Key Laboratory of Quantum Information and Quantum Optoelectronic Devices, Xi'an Jiaotong University, Xi'an 710049, Shaanxi, China}

\author{Chenwei Lv}
\email{lvc@purdue.edu}
\affiliation{Department of Physics and Astronomy, Purdue University, West Lafayette, IN, 47907, USA}

\author{Shengli Ma}
\email{slma1987@xjtu.edu.cn}
\affiliation{School of Physics, Xi'an Jiaotong University, Xi'an 710049, Shaanxi, China}
\affiliation{Shaanxi Province Key Laboratory of Quantum Information and Quantum Optoelectronic Devices, Xi'an Jiaotong University, Xi'an 710049, Shaanxi, China}

\author{Ren Zhang}
\email{renzhang@xjtu.edu.cn}
\affiliation{School of Physics, Xi'an Jiaotong University, Xi'an 710049, Shaanxi, China}
\affiliation{Shaanxi Province Key Laboratory of Quantum Information and Quantum Optoelectronic Devices, Xi'an Jiaotong University, Xi'an 710049, Shaanxi, China}

\date{\today}

\begin{abstract}
Motivated by recent advances in quantum dynamics, we investigate the dynamics of the system with $SU(1,1)$ symmetry. Instead of performing the time-ordered integral for the evolution operator of the time-dependent Hamiltonian, we show that the time evolution operator can be expressed as an $SU(1,1)$ group element. Since the $SU(1,1)$ group describes the ``rotation'' on a hyperbolic surface, the dynamics can be visualized on a Poincar\'e disk, a stereographic projection of the upper hyperboloid. As an example, we present the trajectory of the revival of Bose-Einstein condensation and that of the scale-invariant Fermi gas on the Poincar\'e disk. Further considering the quantum gas in the oscillating lattice, we also study the dynamics of the system with time-dependent single-particle dispersion. Our results are hopefully to be checked in current experiments.
\end{abstract}

\maketitle

\section{Introduction}
The study of quantum dynamics is attracting more and more both experimental and theoretical efforts in cold atom physics. One of the prominent advantages of cold atom physics is flexible manipulation. Both the single-particle Hamiltonian and the pairwise interaction can be precisely controlled by external fields~\cite{optical-lattice,Feshbach}. Using the optical lattice and Feshbach resonance, one can tune the single-particle dispersion and the interaction strength, respectively. As a result, the quantum dynamics in cold atomic gases can be studied in a controllable manner.

Recently, the experimental group from Chicago discovered a series of intriguing dynamics in Bose gases by implementing an oscillating interaction, such as the Bose fireworks~\cite{fireworks} and the revival of Bose-Einstein condensation (BEC)~\cite{BECrevivel,revival2}.
In a proper external potential, the spatial configuration of Bose gas confined in quasi-two-dimensional can be manipulated as well. For instance, the box potential~\cite{box-trap,box-trap2}, circle potential~\cite{circle-trap}, and triangle potential~\cite{ENStriangle-potential} have been realized. The experimental group from ENS observes that BEC in the circle and triangle potential will revive after an appropriate time even the potential has been removed~\cite{ENStriangle-potential}. This phenomenon is quite counterintuitive as the initial state of a many-body system usually does not revive. For Fermi gas in a harmonic trapping potential with unitary or without interaction, the system size shows a discrete scaling law as the trapping frequency changes in a proper way, which is also named as ``Efimovian expansion''~\cite{efimovexp,efimovexp2,dynamicEfimov}. These experiments are quite different from each other in the aspect of statistics, dimensionality, and extra confinement. However, they share the common feature that the dynamics can be described by the $SU(1,1)$ group.

It is instructive to compare the $SU(1,1)$ and $SU(2)$ group~\cite{SU2SU11}. Similar to its $SU(2)$ counterpart, the $SU(1,1)$ group is generated by three operators $\hat{K}_{0}, \hat{K}_{1}$ and $\hat{K}_{2}$ which satisfy the commutation relation $[\hat{K}_{1},\hat{K}_{2}]=-i\hat{K}_{0}, [\hat{K}_{0},\hat{K}_{1}]=i\hat{K}_{2}$ and $[\hat{K}_{2},\hat{K}_{0}]=i\hat{K}_{1}$~\cite{Formula,SO21}. It is known that $SU(2)$ group can describe the rotation on a Bloch sphere, while the $SU(1,1)$ group describes the ``rotation'' on a hyperbolic surface~\cite{SU2group,SU11rotation}. The reason we use the quotation marks is following. Using the commutation relation, one could readily prove that $\hat{K}_{0}$ indeed generates the usual rotation around a particular axis, say, the $z$-axis. However, $\hat{K}_{1}$ and $\hat{K}_{2}$ describe the pseudo-rotation (or boost) along the $y$- and $x$-axis, respectively. To be specific,
\begin{align}
\label{rotation-exp}
&e^{-i\theta \hat{K}_{0}}\hat{K}_{1(2)}e^{i\theta \hat{K}_{0}}=\hat{K}_{1(2)}\cos\theta +(-)\hat{K}_{2(1)}\sin\theta;\\
&e^{-i\theta \hat{K}_{1}}\hat{K}_{2(0)}e^{i\theta \hat{K}_{1}}=\hat{K}_{2(0)}\cosh\theta -\hat{K}_{0(2)}\sinh\theta;\\
&e^{-i\theta \hat{K}_{2}}\hat{K}_{0(1)}e^{i\theta \hat{K}_{2}}=\hat{K}_{0(1)}\cosh\theta +\hat{K}_{1(0)}\sinh\theta,
\end{align}
where $\theta$ is the (pseudo-)rotational angle~\cite{Formula}.
With the aid of this geometric picture, the quantum dynamics can be visualized on the Poincar\'e disk, a stereographic projection of upper hyperboloid, which provides a straightforward intuition of the quantum dynamics of a many-body system. The studies on the dynamics of BEC and the breathing mode in quantum gas are underpinned by the $SU(1,1)$ group~\cite{Breathing-mode0,Breathing-mode,Breathing-mode2,Breathing-mode3,Breathing-mode4,Breathing-mode5,Breathing-mode6,Breathing-mode7,Many-body-echo,Cheng,periodically-Driven} and the corresponding geometric visualization~\cite{GeometryBEC,SU11echo}. As such, it is natural to generalize the geometric visualization to more systems, the dynamics of which are governed by the $SU(1,1)$ group.

Another benefit of the $SU(1,1)$ group is the simplification of the calculation for the quantum dynamics, even when the Hamiltonian is time-dependent~\cite{SU11Coherent}. The conventional wisdom for evaluating the evolution is to perform the time-ordered integral, which is numerically time-consuming, and the analytical treatment is only available in the case that the interaction is perturbative. In contrast, when the Hamiltonian is expressed as the linear combination of the $SU(1,1)$ generators, the evolution operator becomes an $SU(1,1)$ group element. As such, the evolution of the initial state is obtained analytically, which enables us to investigate more generic quantum dynamics.

As of now, the dynamics induced by the manipulation of the pairwise interaction and the external potential have been intensively studied. A natural question arises. What happens if the single-particle dispersion is time-dependent? This can be achieved by considering cold atoms in a periodic driven optical lattice, the depth of which is oscillating. For a tight-binding model, the bandwidth becomes time-dependent. If only low-energy physics is considered, a time-dependent effective mass will capture the dynamics~\cite{Band-structure-effmass}.

The arrangement of this manuscript is as follows. In Section~\ref{sec:generalformula}, we present the general formalism for the quantum dynamics with $SU(1,1)$ symmetry. In Section~\ref{sec:revisit}, the geometric visualization for BEC and scale-invariant Fermi gases are shown. Then, the quantum dynamics for quantum gas in an oscillating optical lattice is studied in Section ~\ref{sec:opticallattice}. At last, we summary our results in Section~\ref{sec:conclusion}.

\section{General formulism}
\label{sec:generalformula}
In this manuscript, the Hamiltonian we consider is of the following generic form,
\begin{align}
\label{generic-H}
\hat{H}(t)=\alpha(t)\hat{K_{0}}+\beta(t)\hat{K}_{1},
\end{align}
where $\alpha(t)$ and $\beta(t)$ are time-dependent. Here we would like to state that the absence of $\hat{K}_{2}$ does not affect the generality of the Hamiltonian in Eq.~(\ref{generic-H}). Suppose that there exists an extra term $\gamma(t)K_{2}$ in the Hamiltonian, one could always perform a rotation 
$e^{i\varphi (t)\hat{K}_{0}}$
to remove the $\hat{K}_{2}$ term, where 
$\varphi (t)${$=-\arctan \left(\gamma (t)/\beta (t)\right) $}.
For different systems, the definition of $\hat{K}_{i}$ varies. For studying the dynamics, the most direct approach is to evaluate the evolution operator $\hat{U}(t,0)={\cal T}\exp\left(-i\int_{0}^{t}\hat{H}(\tau)d\tau\right)$, where ${\cal T}$ denotes the time-ordering operator. However, this method becomes formidable in the many-body system, due to the exponentially increase of the dimension of the Hilbert space. Since the commutation relation is closed, the evolution operator can be written as~\cite{SU11Coherent}
\begin{align}
\label{evolution-operator}
\hat{U}(t,0)=e^{\zeta_{+}(t) \hat{K}_{+}}e^{\hat{K}_{0}\ln\eta(t)}e^{\zeta_{-}(t) \hat{K}_{-}},
\end{align}
where $\zeta_{\pm}(t)$ and $\eta(t)$ are functions of $\alpha(t), \beta(t)$, $\hat{K}_{\pm}\equiv\hat{K}_{1}\pm i\hat{K}_{2}$ are the ladder operators. To be specific, 
we define $|k,j\rangle $ as the common eigenstates of $\hat{K}_{0}$ and the Casimir operator $\hat{C}=\hat{K}_{0}^{2}-\hat{K}_{1}^{2}-\hat{K}_{2}^{2}$, that is, $\hat{K}_{0}|k,j\rangle=j|k,j\rangle $ and $\hat{C}|k,j\rangle =k(k-1)|k,j\rangle $. It can be proved that $\hat{K}_{\pm }|k,j\rangle \propto |k,j\pm 1\rangle $. The requirement that $\langle k,j|\hat{K}_{\mp }\hat{K}_{\pm }|k,j\rangle \geq 0$, i.e., $j(j\pm 1)-k(k-1)\geq 0$ indicates that $j$ is bounded from below with $j_{\min }=k$.
The Hilbert space spanned by $\{|k,k\rangle,\hat{K}_{+}|k,k\rangle,\hat{K}^{2}_{+}|k,k\rangle,...\}$ constitutes a representation of the $SU(1,1)$ group. In this manuscript, we focus on this representation and denote the ground state of $\hat{K}_{0}$ as $|k,k\rangle$.

To determine the coefficients $\zeta_{\pm}(t)$ and $\eta(t)$, on the one hand, we solve the differential equation of the evolution operator $i\partial_{t}\hat{U}(t,0)=\hat{H}(t)\hat{U}(t,0)$,
which can be recasted as
\begin{align}
\label{differentialU1}
\frac{\partial_{t}\hat{U}(t,0)}{\hat{U}(t,0)}=-i\alpha(t)\hat{K_{0}}-\frac{i}{2}\beta(t)(\hat{K}_{+}+\hat{K}_{-}).
\end{align}
This approach is straightforward but formidable in the many-body system.
On the other hand, it is straightforward to prove the following equation from Eq.~(\ref{evolution-operator}),
\begin{align}
\frac{\partial _{t}\hat{U}(t,0)}{\hat{U}(t,0)}& =\frac{1}{\eta (t)}\left[
\frac{\partial \eta (t)}{\partial t}-2\zeta _{+}(t)\frac{\partial \zeta
_{-}(t)}{\partial t}\right] \hat{K_{0}}  \notag  \label{differentialU2} \\
& +\left[ \frac{\partial \zeta _{+}(t)}{\partial t}-\frac{\zeta _{+}(t)}{%
\eta (t)}\frac{\partial \eta (t)}{\partial t}+\frac{\zeta _{+}(t)^{2}}{\eta
(t)}\frac{\partial \zeta _{-}(t)}{\partial t}\right] \hat{K}_{+}  \notag \\
& +\frac{1}{\eta (t)}\frac{\partial \zeta _{-}(t)}{\partial t}\hat{K}_{-}.
\end{align}
For the detailed derivation, please refer to Appendix \ref{sec:app1}.
Comparing the right hand side of Eq.~(\ref{differentialU1}) and that of Eq.~(\ref{differentialU2}), we immediately obtain the following equation set,
\begin{align}
\label{equationset1}
-i\alpha(t)&=\frac{1}{\eta(t)}\left[\frac{\partial \eta(t)}{\partial t}-2\zeta_{+}(t)\frac{\partial \zeta_{-}(t)}{\partial t}\right];\\
\label{equationset2}
-\frac{i}{2}\beta(t)&=\frac{\partial \zeta_{+}(t)}{\partial t}-\frac{\zeta_{+}(t)}{\eta(t)}\frac{\partial \eta(t)}{\partial t}+\frac{\zeta_{+}(t)^{2}}{\eta(t)}\frac{\partial \zeta_{-}(t)}{\partial t};\\
\label{equationset3}
-\frac{i}{2}\beta(t)&=\frac{1}{\eta(t)}\frac{\partial \zeta_{-}(t)}{\partial t}.
\end{align}
This equation set plays a central role in the study of quantum dynamics. In the following sections, we will apply these results to different cases. Solving this algebraic equation under the initial conditions $\zeta_{\pm}(0)=0,\eta(0)=1$, the dynamics of the system is obtained by implementing the evolution operator defined in Eq.~(\ref{evolution-operator}) on the initial state. The initial time is set as $t_{0}=0$. We would like to point out that our approach is valid for any system with the Hamiltonian of the form in Eq.~(\ref{generic-H}), and does not respect the dimensionality, statistics, and interaction strength.

In case that the initial state $|\psi_{i}\rangle=|k,k\rangle$ is the ground state of $\hat{K}_{0}$, $\hat{K}_{-}|\psi_{i}\rangle=0$, the evolution operator can be further simplified as $\hat{U}(t,0)=\eta(t)^{k}e^{\zeta_{+}(t) \hat{K}_{+}}$. The effect of the evolution operator then becomes the same as that of a displacement operator $\hat{D}(\xi(t))=\exp\left(\xi(t) \hat{K}_{+}-\xi^{*}(t)\hat{K}_{-}\right)$ with $\zeta_{+}(t)=\frac{\xi(t)}{|\xi(t)|}\tanh|\xi(t)|$. Then the coherent state of $SU(1,1)$ is defined as~\cite{SU2group}
\begin{align}
\label{coherenstate}
|\zeta_{+}(t),k\rangle=\left(1-|\zeta_{+}(t)|^{2}\right)^{k}\sum_{n=0}^{\infty}\frac{1}{n!}\zeta_{+}^{n}(t)\hat{K}_{+}^{n}|k,k\rangle,
\end{align}
which can be represented by a point on the Poincar\'e disk~\cite{GeometryBEC,SU11echo}. As a result, the evolution of a series system with $SU(1,1)$ symmetry can be visualized on the Poincar\'e disk. In the next section, we take two examples to demonstrate this visualization.

\begin{figure*}[tbp]
\centering
 \includegraphics [width=0.95\textwidth]{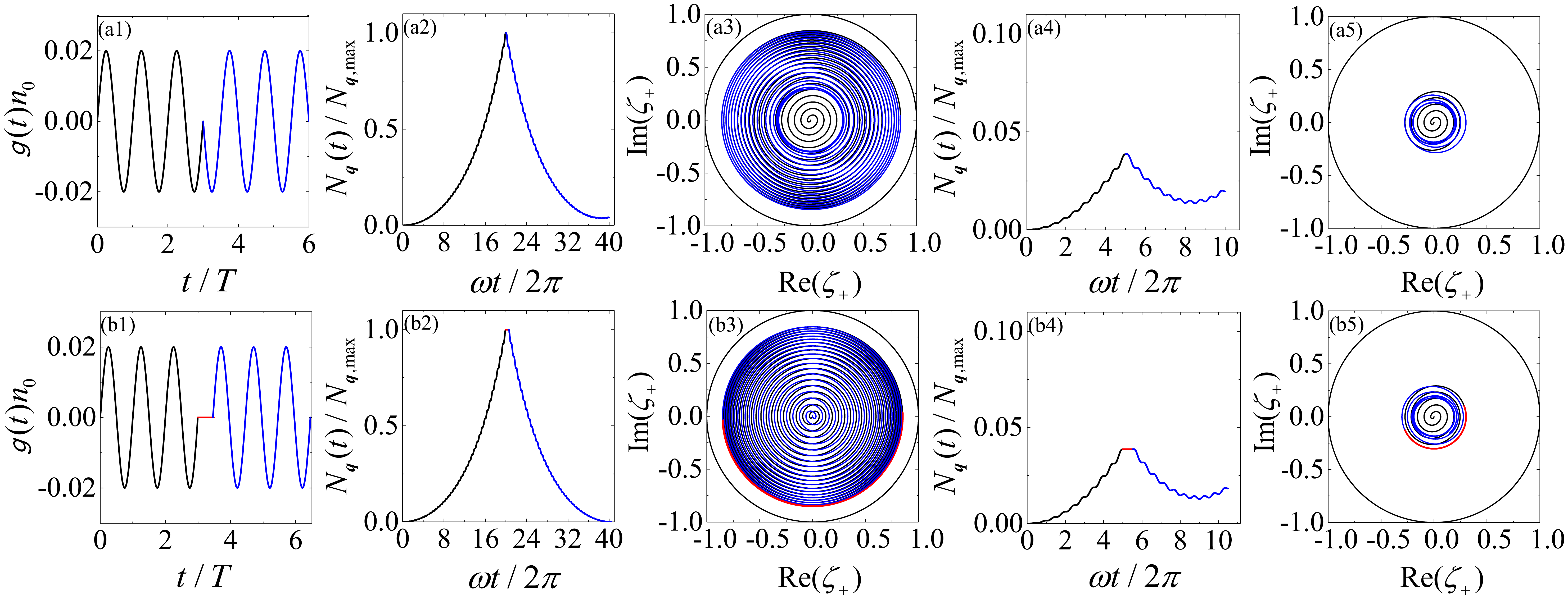}
 \caption{Quantum dynamics of Bose-Einstein condensation under the oscillating $s$-wave scattering length. Oscillation schemes (I) and (II) are schematically shown in (a1) and (b1), respectively. (a2) and (b2): The respective resonant mode particle number as a function of time. 
 Here we define the resonant mode is $\hbar ^{2}q_{\mathrm{res}}^{2}/(2m)=\left( 1/2+3\Delta ^{2}/8\right) \hbar
\omega $ with $\protect\omega$ being the oscillating
frequency.
 The non-monotonic profile indicates BEC revives, and the revival is perfect under the scheme (II). (a3) and (b3) are the visualization of the quantum dynamics on Poincar\'e disk. For both schemes, the start point locates at the disk center. However, the ending point for scheme (I) is elsewhere but the disk center and the ending point for scheme (II) is also the disk center. The similar calculation for the non-resonant mode are shown in (a4,a5,b4,b5). In our calculation, the momentum of the non-resonant mode is $q=0.99q_{\rm res}$, and the interaction strength is
 $g_{0}n_{0}=0.02$. $N_{{\bf q},\rm max}$ is the maximum of $N_{\bf q}$ for the resonant mode.\label{fig:fig1}}
\end{figure*}

\section{Quantum dynamics of Bose-Einstein condensation and scale-invariant Fermi gas}
\label{sec:revisit}
In this section, we implement the approach described in Section~\ref{sec:generalformula} to the bosonic and fermionic system. The revival of BEC and Efimovian expansion in the scale-invariant Fermi gas can be readily obtained, and evolutionary trajectories of the systems are shown on the Poincar\'e disk.
\subsection{Revival of BEC}
We start with the revival of BEC with the oscillating scattering length. In Ref.~\cite{BECrevivel}, the authors observed that the nonzero momentum particle number $N_{\bf q}$ exponentially increases when the $s$-wave scattering length $a_{s}$ is sinusoidally varying. After some time, the phase of $a_{s}$ suddenly changes by $\pi$, and $N_{\bf q}$ begins to decrease, that is, BEC begins to revive. The following theoretical work proposed a more efficient revival protocol by turning off the interaction for a while but keeping the phase invariant, which is named as the ``Many-Body Echo''~\cite{Many-body-echo}. The $SU(1,1)$ echo, a counterpart of the $SU(2)$ spin echo~\cite{spinecho1,spinecho2}, has been proposed and explains the experimental observation~\cite{SU11echo}. The Bogoliubov Hamiltonian is written as $\hat{H}_{\rm Bog}=\frac{g(t)N^{2}}{2V}+\sum_{{\bf q}\neq0}\left(\hat{H}_{\rm Bog}^{\bf q}(t)-\frac{\epsilon_{\bf q}+g(t)n_{0}}{2}\right)$ with
\begin{align}
\label{bogoliubyK}
\hat{H}_{\mathrm{Bog}}^{\mathbf{q}}(t)& =\frac{\left( \epsilon _{\mathbf{q}%
}+g(t)n_{0}\right) }{2}\left( {\hat{a}_{\mathbf{q}}^{\dagger }\hat{a}_{%
\mathbf{q}}}+{\hat{a}_{-\mathbf{q}}\hat{a}_{-\mathbf{q}}^{\dagger }}\right)\notag\\
& +\frac{g(t)n_{0}}{2}\left( {\hat{a}}_{\mathbf{q}}^{\dagger }{\hat{a}}_{-%
\mathbf{q}}^{\dagger }+{\hat{a}}_{\mathbf{q}}{\hat{a}}_{-\mathbf{q}}\right) ,
\end{align}
where 
$\hat{a}_{\mathbf{q}}^{\dagger }$ and $\hat{a}_{\mathbf{q}}$ are the creation and annihilation operator of bosons with momentum ${\bf q}$, respectively. $g(t)=4\pi\hbar^{2} a_{s}(t)/m$ denotes the interaction strength, $\epsilon_{\bf q}=\hbar^{2}q^{2}/(2m)$,
where $m$ is the atomic mass.
$N$ and $V$ are the particle number and system volume, respectively. $n_{0}$ is the zero-momentum particle density. Now we define the generators as follows~\cite{Formula},
\begin{align}
\label{BECgenerator1}
\hat{K}_{0}&=\frac{1}{2}\left(\hat{a}_{\bf q}^{\dagger}\hat{a}_{\bf q}+\hat{a}_{-\bf q}\hat{a}_{-\bf q}^{\dagger}\right);\\
\label{BECgenerator2}
\hat{K}_{1}&=\frac{1}{2}\left(\hat{a}_{\bf q}^{\dagger}\hat{a}_{-\bf q}^{\dagger}+\hat{a}_{\bf q}\hat{a}_{-\bf q}\right);\\
\label{BECgenerator3}
\hat{K}_{2}&=\frac{1}{2i}\left(\hat{a}_{\bf q}^{\dagger}\hat{a}_{-\bf q}^{\dagger}-\hat{a}_{\bf q}\hat{a}_{-\bf q}\right),
\end{align}
and the Casimir operator is written as $\hat{C}=\frac{1}{4}(\hat{a}_{\bf q}^{\dagger}\hat{a}_{\bf q}-\hat{a}_{-\bf q}^{\dagger}\hat{a}_{-\bf q})-\frac{1}{4}$. In our case, $\hat{a}_{\bf q}^{\dagger}\hat{a}_{\bf q}=\hat{a}_{-\bf q}^{\dagger}\hat{a}_{-\bf q}$, and the good quantum number associated with $\hat{C}$ is $k=1/2$. With the aid of Eqs.(\ref{BECgenerator1}-\ref{BECgenerator3}), the Bogoliubov Hamiltonian can be recasted as
\begin{align}
\hat{H}_{\mathrm{Bog}}^{\mathbf{q}}(t)=2\left[ \epsilon _{\mathbf{q}%
}+g(t)n_{0}\right] \hat{K}_{0}+g(t)n_{0}\left( \hat{K}_{+}+\hat{K}%
_{-}\right) .
\end{align}
As such, the approach presented in Section ~\ref{sec:generalformula} is applicable to the dynamics of BEC, and $\alpha(t)=2\left[\epsilon_{\bf q}+g(t)n_{0}\right]$, 
$\beta(t)=2g(t)n_{0}$.
Substituting the expression of $\alpha(t)$ and $\beta(t)$ into Eqs.~(\ref{equationset1}-\ref{equationset3}) and using the initial condition $\zeta_{\pm}(0)=0,\eta(0)=1$, we shall obtain the evolution operator at any time.

Considering that the initial state in the experiment is a condensate at the zero momentum, we could geometrize the dynamics on Poincar\'e disk. Specifically, the initial state can be written as $|\psi_{i}\rangle=\left(\hat{a}^{\dagger}_{{\bf q}=0}\right)^{N}|0\rangle$ with $N$ being the particle number. Then the time-dependent wave function becomes the coherent state
\begin{align}
\label{BECwf}
|\psi(t)\rangle=|\zeta_{+}(t),\frac{1}{2}\rangle=\left(1-|\zeta_{+}(t)|^{2}\right)^{\frac{1}{2}}e^{\zeta_{+}(t) \hat{K}_{+}}|\psi_{i}\rangle,
\end{align}
which can be represented by a point on the Poincar\'e disk. As time goes by, the BEC is excited and the nonzero momentum particle number $\hat{N}_{\bf q}=\hat{K}_{0}-\frac{1}{2}$ can be obtained straightforwardly
\begin{align}
\label{kparticlenumber}
\langle \psi (t)|\hat{N}_{\mathbf{q}}|\psi (t)\rangle =\frac{1}{2}\left(
\frac{1+|\zeta _{+}(t)|^{2}}{1-|\zeta _{+}(t)|^{2}}\right) -\frac{1}{2}.
\end{align}

In Fig.\ref{fig:fig1}, we present the dynamics of BEC under two different types of time-dependent interaction strength. In the first scheme, the interaction is sinusoidally oscillating with a frequency of $\omega$, and after a few periods, say $nT$ with $T=2\pi/\omega$, the phase of $g(t)$ suddenly changes by $\pi$, as schematically plotted in Fig.~\ref{fig:fig1}(a1). We could define the resonance mode via 
$\hbar ^{2}q_{\mathrm{res}}^{2}/(2m)=\left( 1/2+3\Delta ^{2}/8\right) \hbar
\omega $, where $\Delta ${$=g_{0}n_{0}/$}$\hbar \omega $, and $g_{0}$ is the oscillation amplitude.
The particle number $N_{\bf q}$ at the resonant mode increases exponentially at the beginning. When the phase changes by $\pi$, $N_{\bf q}$ begins to decrease, and at $t=2nT$, it drops to a small but nonzero value, as shown in Fig.~\ref{fig:fig1}(a2). On the contrary, in the second scheme, the interaction is turned off for a period of $T/2$ at $t=nT$, then oscillates again with the same phase, as schematically shown in Fig.~\ref{fig:fig1}(b1). It is found that at $t=(2n+1/2)T$, $N_{\bf q}$ is vanishing small as shown in Fig.~\ref{fig:fig1}(b2), which means that BEC perfectly revives under the latter scheme.

These two schemes can be visualized on the Poincar\'e disk. We start with the ground state of $\hat{K_{0}}$, which means that the initial points are located at the center of the disk. The Hamiltonian in Eq.(\ref{bogoliubyK}) is the linear combination of $\hat{K}_{0}$ and $\hat{K}_{1}$ that generate rotation and boost, respectively. Thus, the points represented the instant state on the disk will rotate around the center of the disk and move outwards to the boundary of the disk, as illustrated by the black curves in Fig.~\ref{fig:fig1}(a3) and (b3). In the first scheme, the points begin to rotate around and move inwards to the center of the disk when the phase changes at $t=nT$. At $t=2nT$, it stops elsewhere but the center of the disk as shown in Fig.~\ref{fig:fig1}(a3), which indicates that the system does not go back to the initial state. On the contrary, in the second scheme, when the interaction is turned off for a period of $T/2$, the point rotates by $\pi$ around the center of the disk, indicated by the red curve in Figs.~\ref{fig:fig1}(b1-b3). Then it begins to rotate around and move towards the center. Finally, it goes back to the disk center, which implies that the system perfectly revives.
For the non-resonant mode, the scheme above does not provide revivals as shown in Figs.~\ref{fig:fig1}(a4,a5,b4,b5). The revival requires a different waiting time when the interaction is turned off~\cite{GeometryBEC,SU11echo}.

\subsection{Quantum dynamics in Scale-Invariant Fermi gas in harmonic trap}
For simplicity, we firstly focus on the two-body problems. Considering two fermions of spin $\uparrow$ and spin $\downarrow$ in harmonic potential with time-dependent trapping frequency, the Hamiltonian is written as
\begin{align}
\hat{H}_{2\rm b}(t)=-\frac{\hbar^{2}}{2m}(\nabla_{{\bf r}_{1}}^{2}+\nabla_{{\bf r}_{2}}^{2})+\frac{1}{2}m\omega(t)^{2}(r_{1}^{2}+r_{2}^{2}),
\end{align}
where $m$ is the atomic mass, and $\omega(t)$ denotes the time-dependent trapping frequency. ${\bf r}_{1(2)}$ is the coordinate. We define the center of mass (CoM) and relative coordinates as ${\bf R}=({\bf r}_{1}+{\bf r}_{2})/2$ and ${\bf r}={\bf r}_{1}-{\bf r}_{2}$, respectively. The CoM Hamiltonian is not affected by the pairwise interaction, and we focus on the relative motion, the wave function of which is written as $\psi({\bf r},t)$. Since the angular momentum is conserved in this system, we could define a function $u(r,t)=r\phi(r,t)$ that solves Schr\"odinger equation $i\hbar\partial_{t}u(r,t)=\hat{H}u(r,t)$ with the time-dependent Hamiltonian
\begin{align}
\label{FG-hamil}
\hat{H}(t)=-\frac{\hbar^{2}}{2\mu}\left[\frac{d^{2}}{dr^{2}}-\frac{\ell(\ell+1)}{r^{2}}\right]+\frac{1}{2}\mu\omega(t)^{2}r^{2},
\end{align}
where $\phi(r,t)=\psi({\bf r},t)/Y_{lm}(\theta,\varphi)$ is the radical wavefunction, $\mu$ is the reduced mass, and $\ell$ is the good quantum number of angular momentum. Now we define a set of generators as follows,
\begin{align}
\label{def-alg1}
\hat{K}_{0}\hbar\omega_0=&-\frac{\hbar^{2}}{4\mu}\left(\frac{\partial^{2}}{\partial r^{2}}-\frac{\ell(\ell+1)}{r^{2}}\right)+\frac{1}{4}\mu\omega_{0}^{2}r^{2};\\
\label{def-alg2}
\hat{K}_{1}\hbar\omega_0=&-\frac{\hbar^{2}}{4\mu}\left(\frac{\partial^{2}}{\partial r^{2}}-\frac{\ell(\ell+1)}{r^{2}}\right)-\frac{1}{4}\mu\omega_{0}^{2}r^{2};\\
\label{def-alg3}
\hat{K}_{2}\hbar\omega_0=&\frac{\hbar\omega_{0}}{2i}\left(r\frac{\partial }{\partial r}+\frac{1}{2}\right),
\end{align}
where $\omega_{0}\equiv\omega(t_{0})$ is the trapping frequency at $t=t_{0}$ with $t_{0}$ indicating the initial time.
It is straightforward to prove that the operators in Eqs.~(\ref{def-alg1}-\ref{def-alg3}) satisfy the commutation relation of the $SU(1,1)$ generators, and the Casimir operator is then written as $\hat{C}=[\ell(\ell+1)/4-3/16]$.
For the $s$-wave case ($\ell=0$), the good quantum number for the Casimir is $k=3/4$ or $k=1/4$ that also is the ground state energy of $\hat{K}_{0}$ in Eqs.(\ref{def-alg1}-\ref{def-alg3}) in the presence of different interaction. To be specific, (I): for $k=3/4$, we need to solve $\hat{K}_{0}u_{\rm fr}(r)=(3/4)u_{\rm fr}(r)$ under the boundary condition $u_{\rm fr}(r\to\infty)\to0$, then we have $u_{\rm fr}(r)\propto re^{-\mu\omega_{0}r^{2}/(2\hbar)}$. As such, in the limit of $r\to0$, $\frac{u_{\rm fr}(r)}{r}\to1+{\cal O}(r^{2})$, which is the boundary condition of the relative wave function for two non-interacting particles. (II): for $k=1/4$, we solve $\hat{K}_{0}u_{\rm int}(r)=(1/4)u_{\rm int}(r)$ under the boundary condition $u_{\rm int}(r\to\infty)\to0$, and obtain $u_{\rm int}(r)\propto e^{-\mu\omega_{0}r^{2}/(2\hbar)}$. As such, in the limit of $r\to0$, $\frac{u_{\rm int}(r)}{r}\to\frac{1}{r}+{\cal O}(r)$. Recalling that the Bethe-Peierls boundary condition reads~\cite{BPC}
\begin{align}
\psi(r_{ij}\to0)\propto\frac{1}{r_{ij}}-\frac{1}{a_{s}},
\end{align}
we immediately realize that case (I) and (II) corresponds to noninteracting ($a_{s}\to0$) and unitary interacting ($a_{s}\to\infty$) Fermi gas in harmonic trap with frequency of $\omega_{0}$, respectively. For both cases, the interaction, that can be simulated by the pseudo-potential $(4\pi\hbar^{2}a_{s}/m)\delta({\bf r})\partial_{r}r$~\cite{Huang-Yang}, is invariant under scale transformation 
$r\to\Omega r$ with $\Omega$ being an arbitrary real number.
As a result, scale-invariant Fermi gas refers to that without interaction or with unitary interaction.

\begin{figure*}
\centering
 \includegraphics [width=0.7\textwidth]{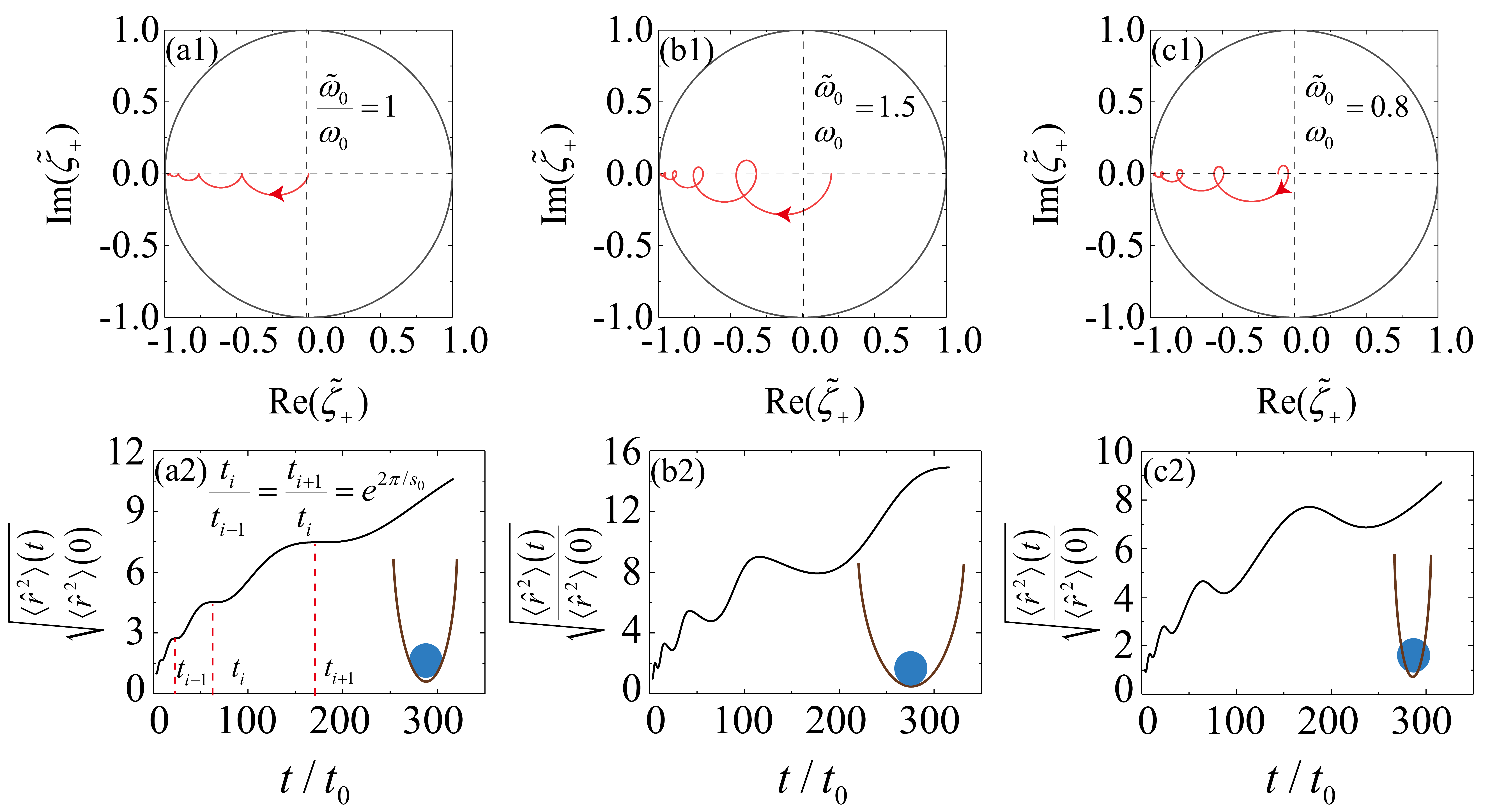}
 \caption{Quantum dynamics of scale-invariant Fermi gas induced by the time-dependent harmonic trap. The frequency varies as $\omega(t)=1/(\sqrt{\lambda}t)$. (a1,a2): The initial state is the ground state of the initial Hamiltonian. On the Poincar\'e disk, the starting point locates at the center of the disk, and $\langle \hat{r}^{2}\rangle(t)$ exhibit a discrete scaling law. (b1,b2): The initial state is in a loosely confined harmonic potential. The starting point locates elsewhere but the center of the disk, and the discrete scaling law disappears. (c1,c2): The initial state is in a tightly confined harmonic potential. Similar to the previous case, the starting point is not the center of the disk and the discrete scaling law disappears. \label{fig:fig2}}
\end{figure*}

Now we consider the quantum dynamics of the scale-invariant Fermi gas. To this end, we firstly recast the Hamiltonian in Eq.(\ref{FG-hamil}) as
\begin{align}
\hat{H}(t)/(\hbar\omega_0)=\left(1+\frac{\omega(t)^{2}}{\omega_{0}^{2}}\right)\hat{K}_{0}+\left(1-\frac{\omega(t)^{2}}{\omega_{0}^{2}}\right)\hat{K}_{1},
\end{align}
then the generic approach for the quantum dynamics in Section \ref{sec:generalformula} is applicable to this system with $\alpha(t)=1+\omega(t)^{2}/\omega_{0}^{2}$ and $\beta(t)=1-\omega(t)^{2}/\omega_{0}^{2}$. Substituting $\alpha(t)$ and $\beta(t)$ into Eqs.(\ref{equationset1}-\ref{equationset3}) and imposing the initial condition $\zeta_{\pm}(0)=0,\eta(0)=1$, we could obtain the evolution operator $\hat{U}(t,0)$. If the initial state were $\phi_{i}(r)=\frac{u_{\rm fr}(r)}{r}$ or $\frac{u_{\rm int}(r)}{r}$, the time-dependent wave function then is written as
\begin{align}
\phi(r,t)=\eta(t)^{\frac{3(1)}{4}}e^{\zeta_{+}(t) \hat{K}_{+}}\phi_{i}(r),
\end{align}
and the quantum dynamics can be visualized on the Poincar\'e disk characterized by $\zeta_{+}$.
For the case that the initial state is not the ground state of $\hat{K}_{0}$, but the ground state of system with trapping frequency $\tilde{\omega}_{0}$, the respective initial state for the noninteracting and unitary gas can be obtained by a dilation
\begin{align}
r\tilde{\phi}_{i}(r)=e^{-i\theta \hat{K}_{2}}u_{\rm fr(int)}(r)
\end{align}
with $\theta=\ln\left(\omega_{0}/\tilde{\omega}_{0}\right)$. Then the time-dependent wave function can be obtained
\begin{align}
\label{tildeoperator}
r\phi(r,t)=&\hat{U}(t,0)e^{-i\theta \hat{K}_{2}}u_{\rm fr(int)}(r)\nonumber\\
=&e^{\tilde{\zeta}_{+}(t) \hat{K}_{+}}e^{\hat{K}_{0}\ln\tilde{\eta}(t)}e^{\tilde{\zeta}_{-}(t) \hat{K}_{-}}u_{\rm fr(int)}(r)\nonumber\\
=&\tilde{\eta}(t)^{\frac{3(1)}{4}}e^{\tilde{\zeta}_{+}(t) \hat{K}_{+}}u_{\rm fr(int)}(r),
\end{align}
where $\hat{U}(t,0)$ is the evolution operator defined in Eq.(\ref{evolution-operator}). To prove the second line, we have resorted to the Baker-Campbell-Hausdorff formula. A straightforward algebra shows that
\begin{align}
\label{tildepara1}
\tilde\zeta_{+}(t)&=\frac{\zeta _+(t)\cosh \frac{\theta}{2}-\left[\eta(t) -\zeta _-(t) \zeta _+(t)\right]\sinh \frac{\theta}{2}}{\cosh \frac{\theta}{2}+\zeta _-(t) \sinh \frac{\theta}{2}};\\
\label{tildepara2}
\tilde\zeta_{-}(t)&=\frac{\zeta _-(t) \cosh\frac{\theta}{2}+\sinh \frac{\theta}{2}}{\cosh \frac{\theta}{2}+\zeta _-(t) \sinh \frac{\theta}{2}};\\
\label{tildepara3}
\sqrt{\tilde\eta(t)}&=\frac{\sqrt{\eta(t)}}{\cosh \frac{\theta}{2}+\zeta _-(t) \sinh \frac{\theta}{2}}.
\end{align}
For details, please refer to Appendix \ref{sec:app2}.
As a result, if the initial state is not the ground state of $\hat{K}_{0}$, the dynamic of scale-invariant Fermi gas will be visualized on the Poincar\'e disk that is characterized by $\tilde{\zeta}_{+}$, instead of $\zeta_{+}$. From Eqs.(\ref{tildepara1}-\ref{tildepara3}), we see that (1): when $\theta=0$, i.e., $\tilde{\omega}_{0}=\omega_{0}$, $\tilde{\zeta}_{\pm}(t)=\zeta_{\pm}(t)$ and $\tilde{\eta}_{\pm}(t)=\eta_{\pm}(t)$; (2): at the initial time $t=0$, $\tilde{\zeta}_{\pm}(0)=\mp\tanh(\theta/2)$ and $\tilde{\eta}(0)=1/\cosh(\theta/2)$. Before further proceeding, an intuitive physical picture will be helpful. $\tilde{\omega}_{0}=\omega_{0}$ means that the initial state is the ground state of the initial harmonic trap, while $\tilde{\omega}_{0}>(<)\omega_{0}$ means that the initial harmonic trap is ``loose (tight)'' for the initial state.

By expressing the time-dependent wavefunction in terms of $SU(1,1)$ coherent state, we can write the expectation value of $\hat{r}^{2}$ as
\begin{align}
\label{r-exceptation}
\langle \hat{r}^{2}\rangle_{\rm fr(int)}(t)&=\left\langle \tilde{\zeta}_{+}(t),\frac{1(3)}{4}\right|\hat{r}^{2}\left| \tilde{\zeta}_{+}(t),\frac{1(3)}{4}\right\rangle\nonumber\\
&=\frac{2}{\mu\omega_{0}^{2}}\left\langle \tilde{\zeta}_{+}(t),\frac{1(3)}{4}\right|\hat{K}_{0}-\hat{K}_{1}\left| \tilde{\zeta}_{+}(t),\frac{1(3)}{4}\right\rangle\nonumber\\
&=\frac{3(1)}{2}\frac{\hbar}{\mu\omega_{0}}\frac{1+|\tilde\zeta_{+}(t)|^{2}-2\Re[\tilde\zeta_{+}(t)]}{1-|\tilde\zeta_{+}(t)|^{2}}.
\end{align}
At the initial time $t=0$, $\langle \hat{r}^{2}\rangle_{\rm fr(int)}(0)=\frac{3(1)}{2}\frac{\hbar}{\mu\tilde\omega_{0}}$ characterizes the cloud size in a harmonic trap of frequency $\tilde\omega_{0}$.
Since the interaction does not affect the CoM motion, the CoM wave function will always be of free form, and the good quantum number of the Casimir operator is $k=3/4$. Following the same derivation, we have
\begin{align}
\label{Rdynamics}
\frac{\langle \hat{R}^{2}\rangle(t)}{\langle \hat{R}^{2}\rangle(0)}=\frac{\tilde{\omega}_{0}}{\omega_{0}}\frac{1+|\tilde\zeta_{+}(t)|^{2}-2\Re[\tilde\zeta_{+}(t)]}{1-|\tilde\zeta_{+}(t)|^{2}}.
\end{align}
It is obvious that the CoM and relative motion have the same dynamics behavior as that of the relative motion. As such, we conclude that the cloud size as a function of time will also obey Eq.(\ref{Rdynamics}). We would like to point out that our approach can be applied to the study of dynamics under arbitrary form of $\omega(t)$ and particle number. The expectation values for observables that are expressed as 
$\hat{K}_{-}^{v}\hat{K}_{0}^{w}\hat{K}_{+}^{p}$ can be readily obtained, where $v,w,p$
are arbitrary non-negative integers~\cite{SU11basis}.

Now we consider a special case that the time-dependent trapping frequency is $\omega(t)=1/(\sqrt{\lambda }t)$ with $\lambda$ being a dimensionless parameter. We denote the initial time as $t_{0}$ and use initial frequency $\omega(t_{0})$ to define the generators in Eqs.(\ref{def-alg1}-\ref{def-alg3}), i.e., $\omega_{0}=\omega(t_{0})$. In Fig.~\ref{fig:fig2}, we illustrate the quantum dynamics starting with different initial Hamiltonian and the visualization of quantum dynamics on the Poincar\'e disk characterized by $\tilde{\zeta}_{+}$. When the initial state is the ground state of the initial harmonic trap, i.e., $\tilde{\omega}_{0}=\omega_{0}$, our results show that the cloud size expands when the trapping frequency becomes smaller. Moreover, a series of plateaus appear for 
$\langle \hat{r}^{2}\rangle (t)$,
the corresponding times of which obey a discrete scaling law $t_{i+1}/t_{i}=e^{2\pi/s_{0}}$ with $s_{0}=\sqrt{4/\lambda-1}$, as shown in Fig.~\ref{fig:fig2}(a). This phenomenon has been observed and named as ``Efimovian expansion''~\cite{efimovexp}. On the Poincar\'e disk, the dynamics trajectory forms a series of similar semicircles, and the initial point locates at the center of the disk. The ending point locates at the boundary of the disk, which implies that the cloud size becomes infinitely large. When the initial state is not the ground state of the initial harmonic trap, that is, the initial harmonic trap is either tight ($\omega_{0}>\tilde{\omega}_{0}$) or loose ($\omega_{0}<\tilde{\omega}_{0}$), the expectation value 
$\langle \hat{r}^{2}\rangle$
grows as time goes by, but on top of which a series of local minima appears instead of plateaus. The discrete scaling law also disappears, as shown in Figs.~\ref{fig:fig2}(b,c). On the Poincar\'e disk, the dynamics trajectory forms a string-like shape. The starting point locates at elsewhere on the horizontal diameter but the center of the disk. This is due to the fact that the starting point for these two cases is related to the center of the disk by a dilation $e^{-i\theta \hat{K}_{2}}$. The ending points also locate at the boundary of the disk, the same as the former case.

\section{Quantum dynamics in oscillating optical lattice}
\label{sec:opticallattice}

In the previous two sections, we have investigated the quantum dynamics induced by the time-dependent interaction strength and the external trapping potential. For cold atoms in optical lattice, the effective mass will be modified due to the distorted dispersion. Considering Bose gas condensed at the $s$-band bottom, we can write its effective mass as
\begin{align}
\label{effectivemass}
m^{*}_{i}=\hbar^{2}\left(\frac{\partial^{2} \epsilon_{\bf q}}{\partial q_{i}^{2}}\right)^{-1}\bigg|_{q_{i}=0},\ i=x,y,z,
\end{align}
where $\epsilon_{\bf q}$ denotes the dispersion. For the tight binding model, we have $\epsilon_{\bf q}=-\sum_{i=x,y,z}t_{i}\cos(q_{i}a)$ with $t_{i}$ being the tunneling strength along the $i$-direction. Since $t_{i}$ is determined by the lattice depth that can be manipulated time-dependently, the dispersion varies for different lattice depth, and so does the effective mass $m^{*}_{i}(t)$~\cite{PDOL}. As such, the Bogoliubov Hamiltonian can be written as
    \begin{align}
    \label{bogoliubyK2}
    \hat{H}_{\mathrm{Bog}}^{\mathbf{q}}(t)& =\frac{1}{2}\left( \frac{\hbar
    ^{2}q^{2}}{2m^{\ast }(t)}+gn_{0}\right) \left( {\hat{a}_{\mathbf{q}%
    }^{\dagger }\hat{a}_{\mathbf{q}}}+{\hat{a}_{-\mathbf{q}}\hat{a}_{-\mathbf{q}%
    }^{\dagger }}\right)  \notag \\
    & +\frac{gn_{0}}{2}\left( {\hat{a}}_{\mathbf{q}}^{\dagger }{\hat{a}}_{-%
    \mathbf{q}}^{\dagger }+{\hat{a}}_{\mathbf{q}}{\hat{a}}_{-\mathbf{q}}\right) ,
    \end{align}
then we define $\alpha(t)=\frac{\hbar^{2}q^{2}}{m^{*}(t)}+2gn_{0}$, and 
$\beta (t)=2gn_{0}$
is a constant. Upon substituting $\alpha(t)$ and $\beta(t)$ into Eqs.~(\ref{equationset1}-\ref{equationset3}), the dynamics can be determined.

\begin{figure}[h]
\centering
 \includegraphics [width=0.4\textwidth]{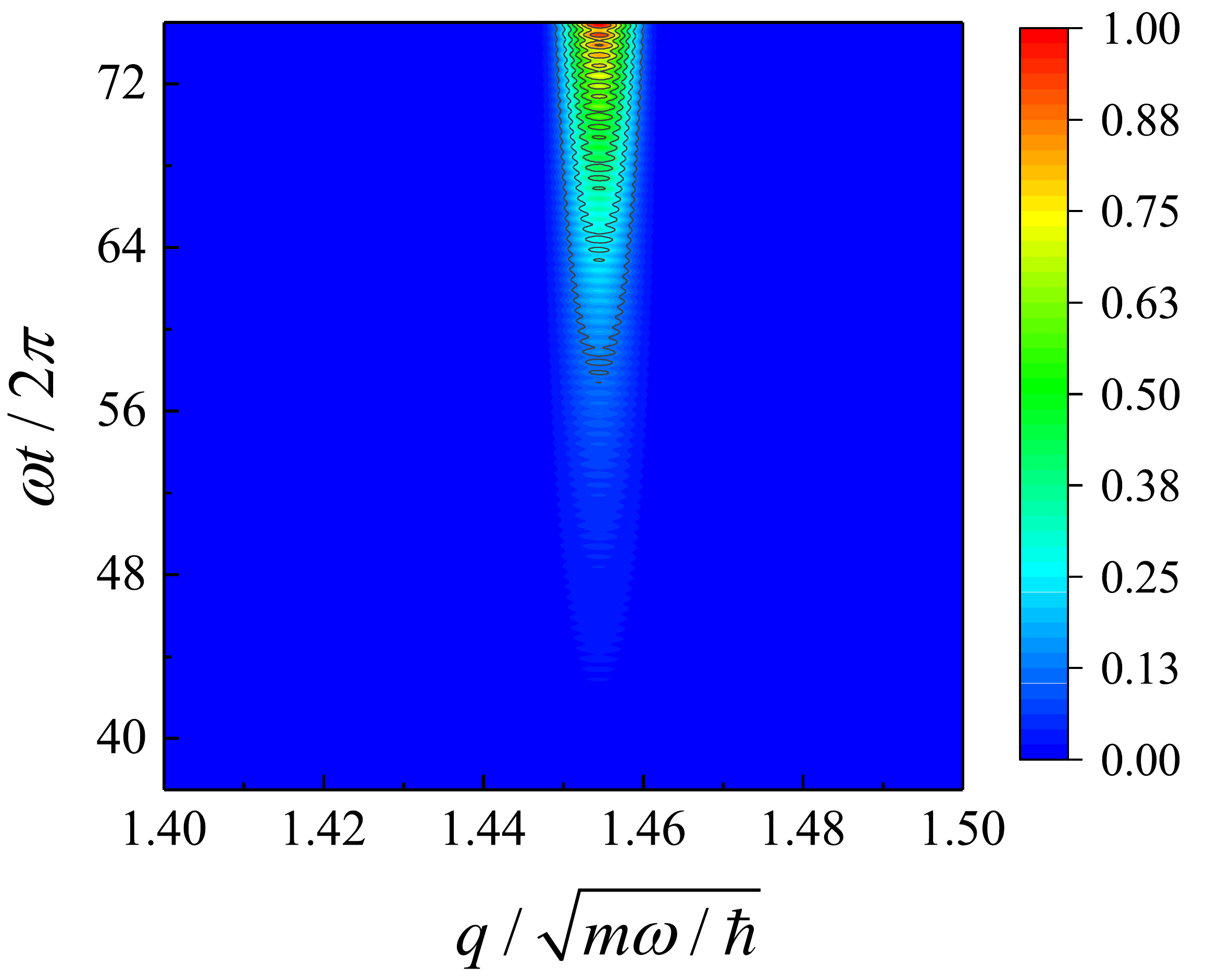}
 \caption{Quantum dynamics of Bose gas in an oscillating optical lattice. Only the particles near the resonant mode are excited. In our calculation, we set $gn_{0}$=0.1.
 \label{fig:fig3}}
\end{figure}

Here we assume that the optical lattice depth varies as $V(t)=V_{0}\sin^{2}(\omega t)$. As such, the kinetic energy can be obtained by exact diagonalization, and varies with a single frequency. Then the time-dependent effective mass can be obtained via Eq.(\ref{effectivemass}).
$V_{0}$ denotes the maximal lattice depth and $E_{R}=$$\hbar ^{2}\pi ^{2}/(2ma^{2})$ is the recoil energy, $m$ is the mass of the particle and $a$ is half the wavelength.
Similar to Eq.(\ref{kparticlenumber}), the particle number of the finite-${\bf q}$ mode can be readily obtained. In Fig.~\ref{fig:fig3}, we present nonzero momentum particle number when time goes by. It is clear that only the particular mode 
$q_{\mathrm{res}}\approx 1.455\sqrt{m\omega /\hbar }$
is excited by the oscillation, where $q_{\rm res}$ is defined as the resonance mode via 
$\hbar^{2}q^{2}_{\rm res}/(2\bar{m}^{*})=n\hbar\omega$.
$\bar{m}^{*}$ denotes the mean value of the varying effective mass, and $n$ is an integer. Here we have shown the case of $n=1$, and the resonant mode with $n\geq2$ can also be observed but with a decreased strength.

\begin{figure}[h]
\centering
 \includegraphics [width=0.4\textwidth]{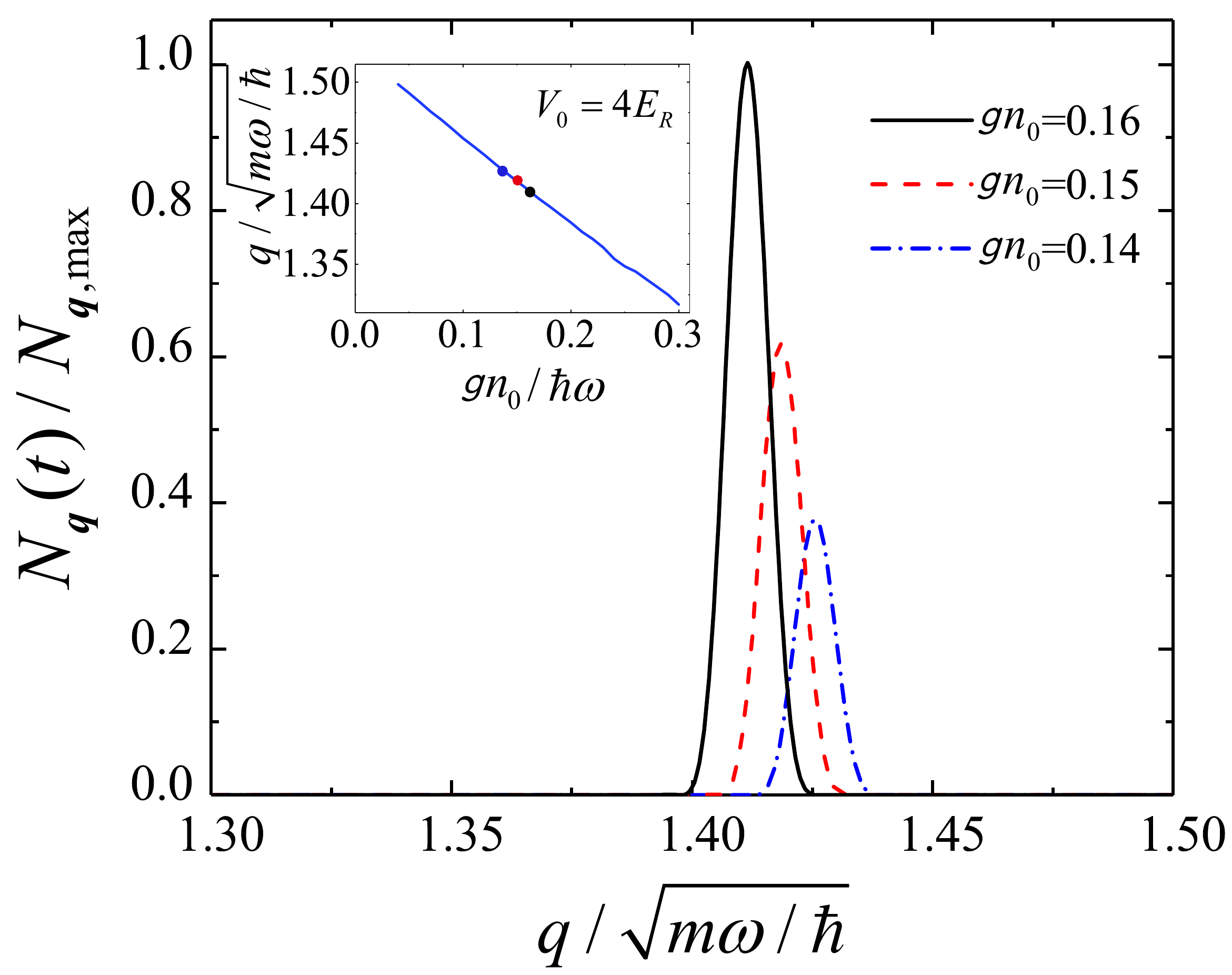}
 \caption{Interaction effect on the resonant mode. The interaction enhances the excitation of the resonant mode. Inset: The interaction shifts the resonant mode toward the lower end. In our calculation, we take shot for the particle number $N_{\bf q}$ at $t=50T$. $V_{0}$ denotes the lattice depth at the beginning and $E_{R}$ is the recoil energy. $N_{{\bf q},{\rm max}}$ is the maximum of particle number for $gn_{0}=0.16$ at $t=50T$. \label{fig:fig4}}
\end{figure}
In addition, we have also studied the interaction effect on the resonant mode. We note that we are constrained in the regime of weak interaction, and the mean-field Bogoliubov theory holds. As illustrated in Fig.~\ref{fig:fig4}, our results show that (1): the stronger the interaction is, the more particles are excited in the resonant mode, which makes sense by referring to Eq.(\ref{bogoliubyK2}). (2): When the interaction strength increases, the resonant mode moves towards the lower end, as shown in the inset. Similar results also exist for the resonant mode with $n\geq2$.

\section{Conclusion \& Outlook}
\label{sec:conclusion}
In summary, we have studied the quantum dynamics of the system with the $SU(1,1)$ symmetry. Instead of performing the time-ordering evolution, the dynamics can be obtained by solving an algebraic equation set. The evolution operator can be expressed as an element of the $SU(1,1)$ group, so that the wave function can be visualized on the Poincar\'e disk, a prototype of the hyperbolic surface. Using Bose and scaling invariant Fermi gas, we demonstrate this approach. On the one hand, our method can readily reproduce the previous results. On the other hand, we also provide an intuitive geometric picture of various quantum dynamical systems. The quantum dynamics of Bose gas in oscillating optical lattice have also been studied.

We would like to point out that our method can also be implemented to the few-body system, such as the study of the dynamics in the three-body problem~\cite{petrov2012fewatom,Efimov2,Chris}. By characterizing how strong the broken of the $SU(1,1)$ symmetry is, our method may also be valid to study the quantum anomaly~\cite{Anomalous0,Anomalous4,Anomalous1,Anomalous2,Anomalous5,Anomalous3}. Since our method does not depend on the configuration of the spatial confinement, it can also be implemented to the study of the parametric excitation in BEC~\cite{Parametric}.

\begin{acknowledgments}
 We are grateful to Qi Zhou and Yangyang Chen for their useful discussions.
 RZ is supported by NSFC (Grant No.12174300) and the National Key R$\&$D Program of China (Grant No. 2018YFA0307601). SLM is supported by the Fundamental Research Funds for the
Central Universities (Grant No. 11913291000022). C.L. is supported by DOE QuantISED program of the theory consortium ``Intersections of QIS and Theoretical Particle Physics" at Fermilab.
\end{acknowledgments}

\appendix
\section{Derivation of Eqs.(\protect\ref{equationset1}-\ref{equationset3})}
\label{sec:app1}
In this appendix, we show the detailed derivation of the Eqs.(\ref{equationset1}-\ref{equationset3}) in the main text. The evolution operator satisfies the following equation
\begin{equation}
\label{appen:evolu1}
\frac{\partial _{t}\hat{U}(t,0)}{\hat{U}(t,0)}=-i\alpha (t)\hat{K_{0}}-\frac{%
i}{2}\beta (t)(\hat{K}_{+}+\hat{K}_{-}).
\end{equation}%
Since the commutation relations of $K_{0,1,2}$ are closed, we can express the evolution operator in the normal-ordered form
\begin{equation}
\hat{U}(t,0)=e^{\zeta _{+}(t)\hat{K}_{+}}e^{\hat{K}_{0}\ln \eta (t)}e^{\zeta
_{-}(t)\hat{K}_{-}},
\end{equation}
which means
\begin{eqnarray}
\partial _{t}\hat{U}(t,0) &=&\frac{\partial \zeta _{+}\left( t\right) }{%
\partial t}\hat{K}_{+}e^{\zeta _{+}(t)\hat{K}_{+}}e^{\hat{K}_{0}\ln \eta
(t)}e^{\zeta _{-}(t)\hat{K}_{-}} \notag \\
&&+\eta \left( t\right) ^{-1}\frac{\partial \eta \left( t\right) }{\partial t%
}e^{\zeta _{+}(t)\hat{K}_{+}}\hat{K}_{0}e^{\hat{K}_{0}\ln \eta (t)}e^{\zeta
_{-}(t)\hat{K}_{-}} \notag \\
&&+\frac{\partial \zeta _{-}\left( t\right) }{\partial t}e^{\zeta _{+}(t)%
\hat{K}_{+}}e^{\hat{K}_{0}\ln \eta (t)}\hat{K}_{-}e^{\zeta _{-}(t)\hat{K}%
_{-}}.
\end{eqnarray}%
For further proceeding, we need to calculate
\begin{eqnarray}
g_{1}(t) &=&e^{\zeta _{+}(t)\hat{K}_{+}}\hat{K_{0}},\ g_{2}(t) =e^{\hat{K}_{0}\ln \eta (t)}\hat{K}_{-}, \\
g_{3}(t) &=&e^{\zeta _{+}(t)\hat{K}_{+}}\hat{K}_{-}.
\end{eqnarray}%
To be specific, it is straightforward to prove the following identity
\begin{eqnarray}
\frac{dg_{1}(t)}{dt} &=&e^{\zeta _{+}(t)\hat{K}_{+}}\hat{K}_{+}\hat{K_{0}} =e^{\zeta _{+}(t)\hat{K}_{+}}(\hat{K_{0}}\hat{K}_{+}-\hat{K}_{+}) \notag
\\
&=&g_{1}(t)\hat{K}_{+}-e^{\zeta _{+}(t)\hat{K}_{+}}\hat{K}_{+}.
\end{eqnarray}%
Under the initial condition that $g_{1}(0)=\hat{K}_{0}$, we find

\begin{equation}
g_{1}(t)=(\hat{K_{0}}-\zeta _{+}\left( t\right) \hat{K}_{+})e^{\zeta _{+}(t)%
\hat{K}_{+}}.
\end{equation}%
Utilizing the same method, we have
\begin{equation}
g_{2}(t)=\eta \left( t\right) ^{-1}\hat{K}_{-}e^{\hat{K}_{0}\ln \eta (t)}.
\end{equation}%
and
\begin{equation}
g_{3}(t)=\left[ \hat{K}_{-}-2\zeta _{+}\left( t\right) \hat{K_{0}}+\zeta
_{+}\left( t\right) ^{2}\hat{K}_{+}\right] e^{\zeta _{+}(t)\hat{K}_{+}}.
\end{equation}
As a result, we have
\begin{align}
&\frac{\partial _{t}\hat{U}(t,0)}{\hat{U}(t,0)} =\frac{\partial \zeta
_{+}\left( t\right) }{\partial t}\hat{K}_{+}+\eta \left( t\right) ^{-1}\frac{%
\partial \eta \left( t\right) }{\partial t}\left[ \hat{K_{0}}-\zeta
_{+}\left( t\right) \hat{K}_{+}\right] \notag \\
&+\eta \left( t\right) ^{-1}\frac{\partial \zeta _{-}\left( t\right) }{%
\partial t}\left[ \hat{K}_{-}-2\zeta _{+}\left( t\right) \hat{K_{0}}+\zeta
_{+}\left( t\right) ^{2}\hat{K}_{+}\right] \notag \\
&=\left[ \frac{\partial \zeta _{+}(t)}{\partial t}-\frac{\zeta _{+}(t)}{%
\eta (t)}\frac{\partial \eta (t)}{\partial t}+\frac{\zeta _{+}(t)^{2}}{\eta
(t)}\frac{\partial \zeta _{-}(t)}{\partial t}\right] \hat{K}_{+} \notag \\
&+\frac{1}{\eta (t)}\left[ \frac{\partial \eta (t)}{\partial t}-2\zeta
_{+}(t)\frac{\partial \zeta _{-}(t)}{\partial t}\right] \hat{K_{0}} \notag
+\frac{1}{\eta (t)}\frac{\partial \zeta _{-}(t)}{\partial t}\hat{K}_{-}.
\end{align}%
Comparing with Eq.(\ref{appen:evolu1}), we obtain Eqs.(\ref{equationset1}-\ref{equationset3}) in the main text,
\begin{eqnarray}
-i\alpha (t) &=&\frac{1}{\eta (t)}\left[ \frac{\partial \eta (t)}{\partial t}%
-2\zeta _{+}(t)\frac{\partial \zeta _{-}(t)}{\partial t}\right] ; \notag \\
-\frac{i}{2}\beta (t) &=&\left[ \frac{\partial \zeta _{+}(t)}{\partial t}-%
\frac{\zeta _{+}(t)}{\eta (t)}\frac{\partial \eta (t)}{\partial t}+\frac{%
\zeta _{+}(t)^{2}}{\eta (t)}\frac{\partial \zeta _{-}(t)}{\partial t}\right]
; \notag \\
-\frac{i}{2}\beta (t) &=&\frac{1}{\eta (t)}\frac{\partial \zeta _{-}(t)}{%
\partial t}.
\end{eqnarray}

\section{Derivation of Eqs.(\protect\ref{tildepara1}-\ref{tildepara3})}
\label{sec:app2}
In this part, we present a detailed derivation for Eqs.(\ref{tildepara1}-\ref{tildepara3}) in the main text.
To this end, we choose the non-unitary representation of the $SU(1,1)$ group. Specifically, $\hat{K}_{0}=\hat{\sigma} _{0}/2$ and
$\hat{K}_{\pm }=i\hat{\sigma} _{\pm }=i(\hat{\sigma} _{1}\pm i\hat{\sigma} _{2})/2$. $\sigma_{0,1,2}$ are Pauli matrices. It can be readily checked that this definition satisfies the communation relation of the $SU(1,1)$ algebra.
Then the operator in Eq.(\ref{tildeoperator}) can be rewritten as
\begin{widetext}
\begin{align}
\label{tilde1}
\hat{U}(t,0)e^{-i\theta \hat{K}_{2}} &=e^{\zeta _{+}(t)\hat{K}_{+}}e^{\hat{K}_{0}\ln \eta (t)}e^{\zeta _{-}(t)%
\hat{K}_{-}}e^{-i\theta \hat{K}_{2}} =e^{\zeta _{+}(t)i\hat{\sigma} _{+}}e^{\ln \eta (t)\hat{\sigma} _{0}/2}e^{\zeta
_{-}(t)i\hat{\sigma} _{-}}e^{\theta \hat{\sigma} _{2}/2} \notag \\
& =\frac{1+\hat{\sigma} _{0}}{2}\left[ \frac{\left[ \eta (t)-\zeta _{-}(t)\zeta
_{+}(t)\right] \cosh \frac{\theta }{2}-\zeta _{+}(t)\sinh \frac{\theta }{2}}{%
\sqrt{\eta (t)}}\right] +\frac{1-\hat{\sigma} _{0}}{2}\left[ \frac{\cosh \frac{\theta }{2}+\zeta
_{-}(t)\sinh \frac{\theta }{2}}{\sqrt{\eta (t)}}\right]  \notag \\
& +i\hat{\sigma} _{+}\left[ \frac{\zeta _{+}(t)\cosh \frac{\theta }{2}-\left[ \eta
(t)-\zeta _{-}(t)\zeta _{+}(t)\right] \sinh \frac{\theta }{2}}{\sqrt{\eta (t)%
}}\right] +i\hat{\sigma} _{-}\left[ \frac{\zeta _{-}(t)\cosh \frac{\theta }{2}+\sinh \frac{%
\theta }{2}}{\sqrt{\eta (t)}}\right]
\end{align}
\end{widetext}

On the other hand, using the same representation, we have the following expression
\begin{align}
\label{tilde2}
&e^{\tilde{\zeta}_{+}(t)\hat{K}_{+}}e^{\hat{K}_{0}\ln \tilde{%
\eta}(t)}e^{\tilde{\zeta}_{-}(t)\hat{K}_{-}} =e^{\tilde{\zeta}_{+}(t)i\hat{\sigma} _{+}}e^{\ln \eta (t)\hat{\sigma} _{0}/2}e^{\zeta
_{-}(t)i\hat{\sigma} _{-}} \notag \\
& =\frac{1+\hat{\sigma} _{0}}{2}\frac{\tilde{\eta}(t)-\tilde{\zeta}_{+}(t)\tilde{%
\zeta}_{-}(t)}{\sqrt{\tilde{\eta}(t)}}+\frac{1-\hat{\sigma} _{0}}{2}\frac{1}{\sqrt{%
\tilde{\eta}(t)}} \notag \\
& +\tilde{\zeta}_{+}(t)\tilde{\eta}(t)^{-1/2}i\hat{\sigma} _{+}+\tilde{\zeta}%
_{-}(t)\tilde{\eta}(t)^{-1/2}i\hat{\sigma} _{-} \notag \\
& +\hat{\sigma} _{0}\left[ \sinh \frac{\ln \tilde{\eta}(t)}{2}-\frac{1}{2}\tilde{%
\zeta}_{+}(t)\tilde{\zeta}_{-}(t)\tilde{\eta}(t)^{-1/2}\right]
\end{align}%
As a result, by comparing the respective terms in Eq.(\ref{tilde1}) and Eq.(\ref{tilde2}), we find
\begin{align}
\frac{1}{\sqrt{\tilde{\eta}(t)}}& =\frac{\cosh \frac{\theta }{2}+\zeta
_{-}(t)\sinh \frac{\theta }{2}}{\sqrt{\eta (t)}}; \\
\frac{\tilde{\zeta}_{+}(t)}{\sqrt{\tilde{\eta}(t)}}& =\frac{\zeta
_{+}(t)\cosh \frac{\theta }{2}-\left[ \eta (t)-\zeta _{-}(t)\zeta _{+}(t)%
\right] \sinh \frac{\theta }{2}}{\sqrt{\eta (t)}}; \\
\frac{\tilde{\zeta}_{-}(t)}{\sqrt{\tilde{\eta}(t)}}& =\frac{\zeta
_{-}(t)\cosh \frac{\theta }{2}+\sinh \frac{\theta }{2}}{\sqrt{\eta (t)}}.
\end{align}%
After some straightforward algebra, we obtain Eqs.(\ref{tildepara1}-\ref{tildepara3}) in the main text,
\begin{align}
\tilde{\zeta}_{+}(t)& =\frac{\zeta _{+}(t)\cosh \frac{\theta }{2}-\left[
\eta (t)-\zeta _{-}(t)\zeta _{+}(t)\right] \sinh \frac{\theta }{2}}{\cosh
\frac{\theta }{2}+\zeta _{-}(t)\sinh \frac{\theta }{2}}; \\
\tilde{\zeta}_{-}(t)& =\frac{\zeta _{-}(t)\cosh \frac{\theta }{2}+\sinh
\frac{\theta }{2}}{\cosh \frac{\theta }{2}+\zeta _{-}(t)\sinh \frac{\theta }{%
2}}; \\
\sqrt{\tilde{\eta}(t)}& =\frac{\sqrt{\eta (t)}}{\cosh \frac{\theta }{2}%
+\zeta _{-}(t)\sinh \frac{\theta }{2}}.
\end{align}


%

\end{document}